\documentclass[twocolumn,prd,psfig,showpacs,superscriptaddress]{revtex4-1}
\usepackage{amssymb}
\usepackage{amsfonts}
\usepackage{amsmath}
\usepackage{graphicx}
\usepackage{tensor}
\usepackage{graphics}
\usepackage{mathptmx}
\usepackage{flushend}

\begin{document}

\title{On Gravitational Energy in Conformal Teleparallel Gravity}
\author{J. G. da Silva}\email{juceliags@gmail.com}\author{S. C. Ulhoa}\email{sc.ulhoa@gmail.com}
\affiliation{Instituto de F\' isica, Universidade de Bras\' ilia, 70910-900, Bras\' ilia, DF, Brazil}

\begin{abstract}
The paper deals with the definition of gravitational energy in conformal teleparallel gravity. The total energy is defined by means of the field equations which allow a local conservation law. Then such an expression is analyzed for a homogeneous and isotropic Universe. This model is implemented by the Friedmann-Robertson-Walker (FRW) line element. The energy of the universe in the absence of matter is identified with the dark energy, however it can be expanded for curved models defining such an energy as the difference between the total energy and the energy of the perfect fluid which is the matter field in the FRW model.
\end{abstract}


\maketitle

\section{Introduction}

Due to recent results obtained in the Laser Interferometer Gravitational-Wave Observatory (LIGO)\cite{abbott2016observation}, the discussion about the transport of energy-momentum by gravitational waves can reemerge~\cite{garecki2002gravitational}. In fact such a discussion has its root on the old problem of gravitational energy definition in General Relativity. Early attempts to define an energy-momentum of the gravitational field lead to pseudo-tensorial quantities~\cite{kox1998collected}. Currently the most spread definition for gravitational energy comes from the ADM formulation which uses the decomposition $3+1$ of General Relativity and its constraints~\cite{Arnowitt:1962hi}. It's useful for some purposes but it fails to provide a general description of the energy inside a finite volume. For instance such a feature is very important to analyze the problem of dark energy. It was postulated to explain the accelerated expansion of the Universe~\cite{Perlmutter:1998np,Riess:1998cb,Silvestri:2009hh}. Of course the measurements take place in the observable Universe which is a finite region. The Einstein equation relates geometry/curvature to any kind of energy and matter, thus the efforts to explain dark energy have concentrated in looking for an extra field in the absence of a reliable definition of gravitational energy. Once such a quantity exists it opens the possibility of dark energy actually being an effect of gravitational field.

In the realm of Teleparallelism Equivalent to General Relativity (TEGR) it is possible to define a gravitational energy-momentum tensor~ \cite{maluf2005gravitational} from which the concept of gravitational energy can be derived. Such an expression is independent of the coordinate system and sensitive to the choice of the reference frame. TEGR is dynamically equivalent to General Relativity since both theories share the field equations. It is possible due to relation between a Riemannian geometry and a Weitzenb\"ockian one. Of course the two theories aren't perfectly interchangeable since tensorial expressions for energy-momentum and angular momentum arise only in Teleparallel gravity. Particularly it has been used to calculate the energy of gravitational waves~\cite{Maluf:2008yy} and of an homogeneous-isotropic Universe~\cite{ulhoa2010gravitational}, it was also used to study quantum gravity~\cite{Ulhoa:2014eka}. We suggest \cite{Maluf:2013gaa} and references therein for a great review on the subject. Although its undeniable successes TEGR is unable to deal with all particle physics since it didn't  exhibit conformal theory. Thus in order to generalize such a theory it's necessary to deal with conformal teleparallel gravity~\cite{Maluf:2011kf}.

In the context of a metric theory of gravity a generalization towards conformal symmetry can be established basically in two analogous ways. The first method is to use the geometric aspect of General Relativity, however the Lagrangian density is constructed out of the Weyl tensor instead the Ricci scalar. The other one is to couple the scalar field with General Relativity and require such a symmetry. The conformal gravity has been used to analyze both astrophysical and cosmological aspects, such as the problem of dark energy and dark matter~\cite{2013Entrp..15..162N}. On the other hand an equivalent form of the Weyl tensor in Teleparallel gravity is lacking up to now, thus in order to establish a conformal teleparallel gravity the second method described above should be used. We hope with such a generalization to deal with the problem of dark energy by means the definition of the total energy. Hence, once the energy of the matter field is a well known quantity, the so called dark energy can be understood as the difference between the total energy and the matter field energy.

This article is divided as follows. In section \ref{tegr} the standard Teleparallel gravity is presented. In section \ref{ctg} the conformal teleparallel gravity is introduced and a general form for the total energy is presented. In section \ref{dark} the previously expression is applied to FRW model and an attempt to identify the dark energy is made. Finally in the last section the conclusions are presented.

\section{Telepallelism Equivalent to General Relativity (TEGR)}\label{tegr}

Teleparallel gravity is an alternative theory of gravitation nevertheless it's dynamically equivalent to General Relativity. General Relativity describes the geometry of the space-time in terms of the curvature tensor while TEGR deals with torsion tensor. Thus the dynamical variables of such a theory is the tetrad field instead of the metric tensor. The variables used in Teleparallel gravity allows it to incorporate the arbitrariness of the reference frame. The metric tensor is written in terms of the tetrads as
$\tensor{g}{_{\mu\nu}}=\tensor{e}{_{a\mu}}\tensor{e}{^{a}_{\nu}}$, where latin indices, $a=(0),(1),(2),(3)$, describe tensorial quantities under Lorentz transformations, while greek ones stand for the diffefeomorphic symmetry.

In order to couple Dirac equation with gravitational field it's necessary the introduction of a connection other than the usual Christoffel symbols, the spin connection $\tensor{\omega}{_{\mu a b}}$, which is sensitive to the $SO (3,1)$ group. It's possible to introduce the spin connection by requiring the vanishing of the covariant derivative of the tetrad field. Thus
\begin{eqnarray}\label{invariance}
&&\nabla_{\mu}\tensor{e}{^{a}_{\nu}}=0,\nonumber\\
&&\partial_{\mu}\tensor{e}{^{a}_{\nu}}-\tensor{\Gamma}{^{\lambda}_{\mu\nu}}\tensor{e}{^{a}_{\lambda}}+ \tensor{\omega}{_{\mu}^{a}_{b}}\tensor{e}{^{b}_{\nu}}=0,
\end{eqnarray}
which leads to
\begin{equation}\label{conecaogeral}
\tensor{\Gamma}{^{\lambda}_{\mu\nu}}=\tensor{e}{^{a\lambda}}{e}{^{b}_{\nu}}\tensor{\omega}{_{\mu a b}}+\tensor{e}{^{a\lambda}}\partial_{\mu}\tensor{e}{_{a\nu}}\,.
\end{equation}
Here $\tensor{\Gamma}{^{\lambda}_{\mu\nu}}$ is a connection related to coordinate transformations, its skew-symmetric part, i.e., the torsion tensor is given by
\begin{equation}
\tensor{T}{^{a}_{\mu\nu}}(e,\omega)= \partial_{\mu}\tensor{e}{^{a}_{\nu}}-\partial_{\nu}\tensor{e}{^{a}_{\mu}}+\tensor{\omega}{_{\mu}^{a}_{\nu}}-\tensor{\omega}{_{\nu}^{a}_{\mu}}\,.
\end{equation}

The vanishing of the spin connection leads to
\begin{equation} \label{tensortorsao}
\tensor{T}{^{a}_{\mu\nu}}(e)= \partial_{\mu}\tensor{e}{^{a}_{\nu}}-\partial_{\nu}\tensor{e}{^{a}_{\mu}}.
\end{equation}
Such a requirement is known as the Teleparallel condition. It allows an equivalence between General Relativity and Teleparallel gravity. It's possible because the spin connection is identically written in terms of the Levi-Civita connection (a torsion free connection related to Christoffel symbols) plus the contortion tensor which is constructed out of the torsion tensor above. In this sense it's possible to show that the TEGR Lagrangian density~\cite{Maluf:2013gaa} is given by
\begin{equation}\label{conformal}
\mathcal{L}(\tensor{e}{_{a\mu}})=-ke\tensor{\Sigma}{^{abc}}\tensor{T}{_{abc}}-\mathcal{L}{_{M}}\,,
\end{equation}
where $\mathcal{L}{_{M}}$ is the Lagrangian density for matter fields, $k=\frac{1}{16\pi}$, $e=\sqrt{-g}$ and
\begin{eqnarray}
\tensor{\Sigma}{^{abc}}=\frac{1}{4}\left(\tensor{T}{^{abc}}+\tensor{T}{^{bac}}-\tensor{T}{^{cab}}\right)+ \frac{1}{2}\left(\tensor{\eta}{^{ac}}\tensor{T}{^{b}}-\tensor{\eta}{^{ab}}\tensor{T}{^{c}}\right)\,,\nonumber\\
\end{eqnarray}
with $\tensor{T}{^{a}}=\tensor{T}{^{b}_{b}^{a}}$.

If one performs a variation of the Lagrangian density with respect to the tetrad field, then the field equation reads
\begin{equation}
\partial_{\lambda}\left(e\tensor{\Sigma}{^{a\mu\lambda}}\right)=\frac{1}{4k}e\left(\tensor{\Theta}{^{a\mu}}+\tensor{\tau}{^{a\mu}}\right)\,,
\end{equation}
where
\begin{eqnarray}
\tensor{\tau}{^{a\mu}}=k\left(4\tensor{\Sigma}{^{b\lambda\mu}}\tensor{T}{_{b\lambda}^{a}}-\tensor{e}{^{a\mu}}\tensor{\Sigma}{^{bcd}}\tensor{T}{_{bcd}}\right)
\end{eqnarray}
is the energy-momentum of the gravitational field and $\Theta^{a\mu}=\frac{1}{e}\frac{\delta{\mathcal{L}{_{M}}}}{\delta{e_{a\mu}}}$ is the energy-momentum tensor of matter fields. It can be integrated over a 3D volume to yield the energy-momentum vector which is invariant under coordinate transformations and dependent on the choice of the reference frame. In the next section it'll be defined a similar quantity for conformal teleparallel gravity.

\section{Conformal Teleparallel Gravity}\label{ctg}

In this section the conformal teleparallel gravity will be presented briefly following reference \cite{Maluf:2011kf}, as well as a definition of total energy-momentum in such a context. It should be noted that the the TEGR Lagrangian density is not invariant under conformal transformation
\begin{equation}
\tensor{g}{_{\mu\nu}} \rightarrow \tensor{\tilde{g}}{_{\mu\nu}}= e^{2\theta (x)}\tensor{g}{_{\mu\nu}},
\end{equation}
where $\theta (x)$ is the conformal factor which is an arbitrary function, this induces a transformation in the tetrad field as $\tilde{e}_{a\mu}=e^{\theta (x)}\tensor{e}{_{a\mu}}$ due to the relation between both fields. In order to see that non-invariance let's consider the change in the torsion contractions in the TEGR Lagrangian density under conformal transformation, they are
\begin{eqnarray}
\tensor{\tilde{T}}{^{abc}}\tensor{\tilde{T}}{_{abc}}=e^{-2\theta}\left(\tensor{T}{^{abc}}\tensor{T}{_{abc}}-4\tensor{T}{^{\mu}}\partial_{\mu}{\theta}+ 6 \tensor{g}{^{\mu\nu}}\partial_{\mu}{\theta}\partial_{\nu}{\theta}\right)\,,\nonumber
\end{eqnarray}
\begin{eqnarray}
\tensor{\tilde{T}}{^{abc}}\tensor{\tilde{T}}{_{bac}}= e^{-2\theta}\left(\tensor{T}{^{abc}}\tensor{T}{_{bac}}-2\tensor{T}{^{\mu}}\partial_{\mu}{\theta}+ 3\tensor{g}{^{\mu\nu}}\partial_{\mu}{\theta}\partial_{\nu}{\theta}\right)\,,\nonumber
\end{eqnarray}
and
\begin{eqnarray}
\tensor{\widetilde{T}}{^{a}}\tensor{\widetilde{T}}{_{a}}=\tensor{e}{^{-2\theta}}\left(\tensor{T}{^{a}}\tensor{T}{_{a}}-6\tensor{T}{^{\mu}}\partial_{\mu}{\theta}+9\tensor{g}{^{\mu\nu}}\partial_{\mu}{\theta}\partial_{\nu}{\theta}\right)\,.\nonumber
\end{eqnarray}
Therefore a Lagrangian density invariant under conformal transformation in the context of Teleparallel gravity is given by
\begin{eqnarray}
\mathcal{L}&=&ke\Big[-\phi^{2}\left(\frac{1}{4}\tensor{T}{^{abc}}\tensor{T}{_{abc}}+\frac{1}{2}\tensor{T}{^{abc}}\tensor{T}{_{bac}}-\frac{1}{3}\tensor{T}{^{a}}\tensor{T}{_{a}}\right)+  \nonumber\\
&&k' \tensor{g}{^{\mu\nu}}D_{\mu}\phi D_{\nu}{\phi}\Big]\,,
\end{eqnarray}
where $D_{\mu}$ is the covariant derivative defined as
\begin{equation}
D_{\mu}{\phi}=\left(\partial_{\mu}-\frac{1}{3}\tensor{T}{_{\mu}}\right)\phi\,,
\end{equation}
the quantity $\phi$ is a scalar field, $\tensor{T}{_{\mu}}=\tensor{T}{^{\lambda}}{_{\lambda\mu}}=\tensor{T}{^{a}_{a\mu}}$ and k' is a coupling constant, it has to be established to recover TEGR when $\phi$ is a constant. Hence for $k'=6$, the conformal teleparallel Lagrangian density reads
\begin{eqnarray}
\mathcal{L}(\tensor{e}{_{a\mu}},\phi)&=&ke\big[-\phi^{2}\tensor{\Sigma}{^{abc}}\tensor{T}{_{abc}}+ 6 \tensor{g}{^{\mu\nu}}\partial_{\mu}\phi \partial_{\nu}\phi - \nonumber \\ &4&\tensor{g}{^{\mu\nu}}\phi(\partial_{\mu}\phi)\tensor{T}{_{\nu}}\big]-\mathcal{L}_M\,.
\end{eqnarray}
It should be pointed out that in this article we're looking for an extension of teleparallel gravity, however a constraint over the scalar field is always possible to establish. In fact it is exactly what it was done with the introduction of the dark fluid in (\ref{eliminarrho}) and (\ref{substituindorho}).

The field equations are obtained from the variation of the above Lagrangian density with respect to the scalar field and the tetrad field. The former variation yields
\begin{eqnarray}
\partial_{\nu}(e\tensor{g}{^{\mu\nu}}\partial_{\mu}\phi)+\frac{1}{6}\phi\left[e\tensor{\Sigma}{^{abc}}\tensor{T}{_{abc}}-2\partial_{\mu}(e\tensor{T}{^{\nu}})\right]=\frac{1}{12 k}\frac{\delta{\tensor{L}{_{M}}}}{\delta{\phi}} \nonumber
\end{eqnarray}
where the terms in brackets is the negative of the Ricci scalar. Thus
\begin{equation}
\partial_{\nu}(e\tensor{g}{^{\mu\nu}}\partial_{\mu}\phi)-\frac{1}{6}e\phi R(e)=\frac{1}{12 k}\frac{\delta{\tensor{L}{_{M}}}}{\delta{\phi}}\,,
\end{equation}
where $ \phi \frac{\delta{\tensor{L}{_{M}}}}{\delta{\phi}}= e\Theta$, with $\Theta=\tensor{g}{_{\mu\nu}}\tensor{\Theta}{^{\nu\mu}}$. The variation with respect of the tetrad field results in
\begin{eqnarray}
&&\partial_{\lambda}(e\phi^2\tensor{\Sigma}{^{a\mu\lambda}})-e\phi^2\big(\tensor{\Sigma}{^{b\lambda\mu}}\tensor{T}{_{b\lambda}^{a}}-\frac{1}{4}\tensor{e}{^{a\mu}}\tensor{\Sigma}{^{bcd}}\tensor{T}{_{bcd}}\big) - \nonumber \\&& \frac{3}{2}
e \tensor{e}{^{a\mu}}\tensor{g}{^{\sigma\nu}}\partial_{\sigma}\phi \partial_{\nu}\phi + 3e \tensor{e}{^{a\nu}}\tensor{g}{^{\sigma\mu}}\partial_{\sigma}\phi \partial_{\nu}\phi + \nonumber \\&& e\tensor{e}{^{a\mu}}\tensor{g}{^{\sigma\nu}}\tensor{T}{_{\nu}}\phi(\partial_{\sigma}\phi)-e\phi \tensor{e}{^{a\sigma}}\tensor{g}{^{\mu\nu}}\left(\tensor{T}{_{\nu}}\partial_{\sigma} \phi+\tensor{T}{_{\sigma}}\partial_{\nu}\phi\right) - \nonumber \\&& e\tensor{g}{^{\sigma\nu}}\phi(\partial_{\sigma}\phi)\tensor{T}{^{\mu a}_{\nu}} - \partial_{\rho}\left[e\tensor{g}{^{\sigma\mu}}\phi(\partial_{\sigma}\phi) \tensor{e}{^{a\rho}} \right] +\nonumber \\ && \partial_{\nu}\left[e\tensor{g}{^{\sigma\nu}}\phi(\partial_{\sigma}\phi) \tensor{e}{^{a\mu}} \right]=\frac{1}{4 k}\frac{\delta{\tensor{L}{_{M}}}}{\delta{\tensor{e}{_{a\mu}}}}\,,\label{equacaotetrada}
\end{eqnarray}
where $\frac{\delta{\tensor{L}{_{M}}}}{\delta{\tensor{e}{_{a\mu}}}}= e\tensor{e}{^{a}_{\nu}} \Theta^{\nu\mu}$. We point out that such a tensor is traceless in order to be compatible with the conformal symmetry. Making use of previous definitions such as
\begin{equation}
\tensor{\tau}{^{a\mu}}=k\left(4\tensor{\Sigma}{^{b\lambda\mu}}\tensor{T}{_{b\lambda}^{a}}-\tensor{e}{^{a\mu}}\tensor{\Sigma}{^{bcd}}\tensor{T}{_{bcd}}\right)
\end{equation}
and using the expression
\begin{eqnarray}
\tensor{t}{^{a\mu}}&=&-4k\Big\{-\frac{3}{2}\tensor{e}{^{a\mu}}\tensor{g}{^{\sigma\nu}}(\partial_{\sigma}\phi)(\partial_{\nu}\phi)+ \nonumber \\ &3&\tensor{e}{^{a\nu}}\tensor{g}{^{\sigma\mu}}(\partial_{\sigma}\phi)(\partial_{\nu}\phi)+\tensor{e}{^{a\mu}}\tensor{g}{^{\sigma\nu}}\tensor{T}{_{\nu}}\phi(\partial_{\sigma}\phi)- \nonumber\\ &\tensor{e}{^{a\sigma}}&\tensor{g}{^{\mu\nu}}\phi\big(\tensor{T}{_{\nu}}\partial_{\sigma}\phi+\tensor{T}{_{\sigma}}\partial_{\nu}\phi\big)-\tensor{g}{^{\sigma\nu}}\tensor{T}{^{\mu a}_{\nu}}\phi(\partial_{\sigma}\phi)-  \nonumber \\ &\Big(\frac{1}{e}\Big)&\Big(\partial_{\rho}\big[e\tensor{e}{^{a\rho}}\tensor{g}{^{\sigma\mu}}\phi(\partial_{\sigma}\phi)\big]-\partial_{\nu}\big[e\tensor{e}{^{a\mu}}\tensor{g}{^{\sigma\nu}}\phi ( \partial_{\sigma}\phi )\big]\Big)\Big\} \nonumber \,,
\end{eqnarray}
then the field equation (\ref{equacaotetrada}) is rewritten as
\begin{eqnarray}
\partial_{\lambda}\big(e\phi^2 \tensor{\Sigma}{^{a\mu\lambda}})=\frac{1}{4k}e(\tensor{e}{^{a}_{\nu}}\tensor{\Theta}{^{\mu\nu}}+\phi^2\tensor{\tau}{^{a\mu}}+\tensor{t}{^{a\mu}}\big)\,.
\end{eqnarray}
This equation is in many aspects similar to the Brans-Dicke field equation, however an important feature should be highlighted: similarly to TEGR, equation (\ref{equacaotetrada}) allows a local conservation law from which it's possible to construct the total energy-momentum vector of the theory. Due to the skew-symmetry in the last two indices of $\tensor{\Sigma}{^{a\mu\lambda}}$,
\begin{equation}
\partial_{\mu}\partial_{\lambda}(e\phi^2 \tensor{\Sigma}{^{a\mu\lambda}})\equiv 0\,,
\end{equation}
then
\begin{equation}
\partial_{\mu}\left[e\left(\tensor{e}{^{a}_{\nu}}\tensor{\Theta}{^{\nu\mu}}+\phi^2\tensor{\tau}{^{a\mu}}+\tensor{t}{^{a\mu}}\right) \right]=0\,.
\end{equation}
Therefore the total energy-momentum in the conformal teleparallel gravity is given by
\begin{equation}
P^{a}=\int_{V}{e(\tensor{e}{^{a}_{\nu}} \tensor{\Theta}{^{0\nu}}+\phi^{2} \tensor{\tau}{^{a0}}+\tensor{t}{^{a0}})d^{3}x}\,, \label{gravitationalenergy1}
\end{equation}
which can be rewritten, using equation (\ref{equacaotetrada}), as
\begin{equation}
P^{a}= 4k \oint{e\phi^{2}\tensor{\Sigma}{^{a0i}}dS_{i}}\,.\label{energia}
\end{equation}
This definition of energy-momentum vector shares the same feature of the expression in TEGR concerning the coordinates transformation and the choice of the reference frame. The zero component of the vector defined in equation (\ref{gravitationalenergy1}) is the total energy, here the term total refers to the fact that there are three separated contributions to the energy, one from the matter field, other from the gravitational energy and another one from the conformal factor. In the next section this will be applied to a Universe model.

\section{Dark Energy in Conformal Teleparallel Gravity}\label{dark}

The idea of dark energy arose with the observation that the Universe is at an accelerating expansion rate~\cite{Perlmutter:1998np,Riess:1998cb,Silvestri:2009hh}. The usual approach to deal with that is to look for some matter field that accelerates the Universe, for instance some dark fluid that turns the deceleration parameter negative or an arbitrary field added to the right-side of Einstein equation. Thus it makes sense to follow this path in a theory that don't predict gravitational energy such as General Relativity. However one has to admit that the existence of gravitational energy plays an important role in this subject. In this section we calculate the net energy in the Universe and associate it to the so called dark energy.

In our recent work \cite{silva2016friedmann}, we worked with FRW metric, solving the model for conformal teleparallel gravity. Then we explored how the scale factor and the scalar field could define a dark fluid responsible for the acceleration of the Universe. Now we intend to calculate the actual dark energy, understood here as the difference between the total energy, given by equation (\ref{energia}), and the energy of the perfect fluid, defined by $\Theta^{\mu\nu}$.

The Friedmann-Robertson-Walker line element is given by
\begin{equation}
ds^2=-dt^2+a^2(t)\left[\frac{dr^2}{1-\kappa r^2}+r^2d\theta^2+r^2\sin^2{\theta}d\phi'^2\right],
\end{equation}
where $a(t)$ is scalar factor and $\kappa$ is the space curvature which assumes the values $(-1,0,1)$. It represents a isotropic and homogeneous Universe. For sake of simplicity we take the following tetrad field
 \begin{equation}
 e_{a\mu}= \left(\begin{tabular}{c c c c}
 -1 & 0 & 0 & 0 \\
  0 & $\frac{a(t)}{\sqrt{(1-\kappa r^2)}}$ & 0 & 0 \\
  0 & 0 & $r a(t)$ & 0 \\
 0 & 0 & 0 & $r a(t) \sin{\theta}$ \\
  \end{tabular} \right).
 \end{equation}
Then in order to calculate the total energy it's necessary to have the component $ \tensor{\Sigma}{^{(0)01}} $ which is given by
\begin{equation}
\tensor{\Sigma}{^{(0)01}}=-\frac{1+\kappa r^2}{a(t)^{2} r}
\end{equation}
and the determinant of the tetrad field,
\begin{equation}
e=\frac{a(t)^{3}r^{2}\sin{\theta}}{\sqrt{1-\kappa r^{2}}}\,,\label{determinantofthetetrad}
\end{equation}
thus we integrate over a sphere of radius R and we use $ dS_{1} =d\theta d\phi^{'}$ in the time component of $P^a$, which yields
\begin{eqnarray}
P^{(0)}&=& \lim_{r\rightarrow R} 4k \int_{0}^{2\pi}\int_{0}^{\pi} -a(t)\phi(t)^{2} r\sin{\theta} \sqrt{1-\kappa r^{2}} d\theta d\phi^{'}\nonumber\\
P^{(0)}&=&-a(t)R\phi(t)^{2} \sqrt{1-\kappa R^{2}}\,.\nonumber
\end{eqnarray}
This is the total energy inside a sphere of radius $R$, it's interesting to note that the FRW line element has a dynamical horizon which is given by
\begin{equation*}
R=\frac{1}{\sqrt{\left(\frac{\dot{a}}{a}\right)^2+\frac{\kappa}{a^2}}}\,,
\end{equation*}
hence the total energy $E\equiv P^{(0)}$ of the observable Universe is
\begin{equation}
E=-\frac{a^2\phi^2\left[\dot{a}^2+\kappa\left(1-a^2\right)\right]^{1/2}}{\left(\dot{a}^2+\kappa\right)}\,.\label{energiatotal}
\end{equation}

%\twocolumn{
%
%\begin{figure}
%\includegraphics[scale=0.45]{energiakmenosuma.png}\\
%\includegraphics[scale=0.45]{energiakmenosumb.png}
%\end{figure}
%\begin{figure}
%\includegraphics[scale=0.45]{energiakzeroa.png}\\
%\includegraphics[scale=0.45]{energiakzerob.png}
%\end{figure}
%\begin{figure}
%\includegraphics[scale=0.45]{energiakuma.png}\\
%\includegraphics[scale=0.45]{energiakumb.png}
%\label{fig:3}       % Give a unique label
%\end{figure}
%
%}

The energy of the matter, here a conformal perfect fluid, is calculated as
\begin{equation}
E_m=\int e\tensor{\Theta}{^{(0)0}}d^3x\,, \label{energyofthematter}
\end{equation}
where $\tensor{\Theta}{^{(0)0}}=\rho \phi^2$. Therefore using (\ref{determinantofthetetrad}), equation (\ref{energyofthematter}) becomes
\begin{eqnarray}
E_{m}&=&\int_{0}^{R}\int_{0}^{2\pi}\int_{0}^{\pi}{\frac{a^3r^2\sin{\theta}}{\sqrt{1-\kappa r^2}} \rho \phi^2 d\theta d\phi'dr}\nonumber\\
&=&4\pi a^3\rho\phi^2\int_{0}^{R}{\frac{r^2}{\sqrt{1-\kappa r^2}}dr}\,,\nonumber\label{plotofthematter}
\end{eqnarray}
which with
\begin{equation*}
\int_{0}^{R}{\frac{r^2}{\sqrt{1-\kappa r^2}}dr}=\frac{1}{2\kappa}\left[\frac{\sin^{-1}{\left(\sqrt{\kappa}R\right)}}{\sqrt{\kappa}}-R\sqrt{1-\kappa R^2}\right]
\end{equation*}
yields
\begin{small}
\begin{eqnarray}
E_{m}&=&\frac{2\pi a^3\rho\phi^2}{\kappa^{3/2}}\left[\sin^{-1}{\left(\frac{a\sqrt{\kappa}}{\sqrt{\dot{a}^2+\kappa}}\right)}-\left(\frac{a\sqrt{\kappa\dot{a}^2+\kappa^2(1- a^2)}}{\dot{a}^2+\kappa}\right)\right]\,.\label{massa}\nonumber\\
\end{eqnarray}
\end{small}

In view of expressions (\ref{energiatotal}) and (\ref{massa}) it is possible to define the dark energy as the difference between them, $$E_d=E-E_m\,.$$ Thus dark energy would be what remains when the mass of the Universe vanishes, i. e., it's the vacuum energy.

The field equations from the reference \cite{silva2016friedmann} read,
\begin{equation}
3H^2 + \frac{3k}{a^2}+3\beta^2+6H\beta=8\pi \rho\,, \label{eliminarrho}
\end{equation}
\begin{equation}
-2\dot{H} - 3H^2-\frac{k}{a^2}- 4H\beta-\beta^2-2\dot{\beta}=8\pi p\,, \label{substituindorho}
\end{equation}
where $H=\frac{\dot{a}}{a}$ and $\beta=\frac{\dot{\phi}}{\phi}$. It should be noted that the energy-momentum of matter is taken as the perfect fluid, due to the conformal symmetry the relation $p=\frac{1}{3}\rho$ holds. The above equations are under-determined which is the same problem appearing in the Friedmann equations. In order to avoid this problem we note that the field equations behave as if there is a dark fluid with density $\rho_D=\frac{3\beta^2+6H\beta}{8\pi}$ and pressure $p_D=\frac{- 4H\beta-\beta^2-2\dot{\beta}}{8\pi}$. Hence to determine all functions we suppose a equation of state for the dark fluid as $p_D=w\,\rho_D$, where $w$ is the dark fluid parameter. Then, for $\kappa=0$, the time dependence of $$E_d=-\frac{a\phi^2}{H}\left(1+\frac{4\pi\rho a^2}{3H^2}\right)\,,$$ is given below in figures \ref{fig:1}, \ref{fig:2} and \ref{fig:3}.

\begin{figure}
\includegraphics[scale=0.45]{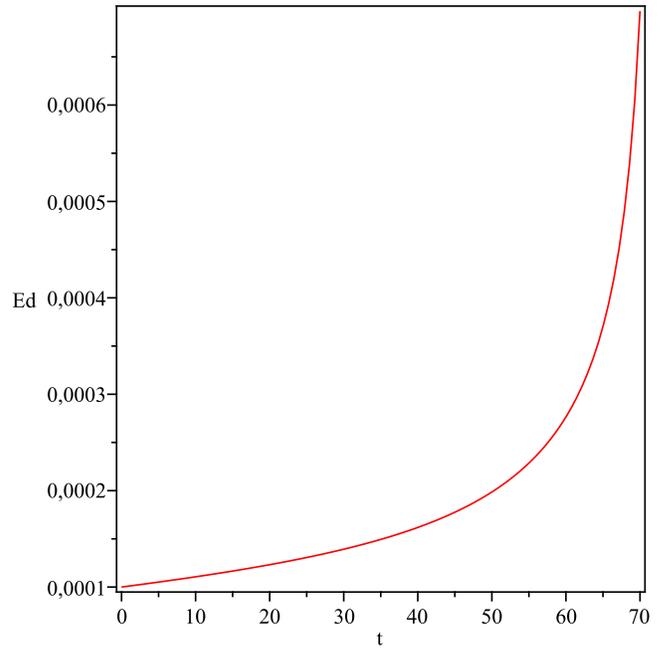}
\caption{$E_d$ for $w=0$.}
\label{fig:1}       % Give a unique label
\end{figure}

\begin{figure}
\includegraphics[scale=0.45]{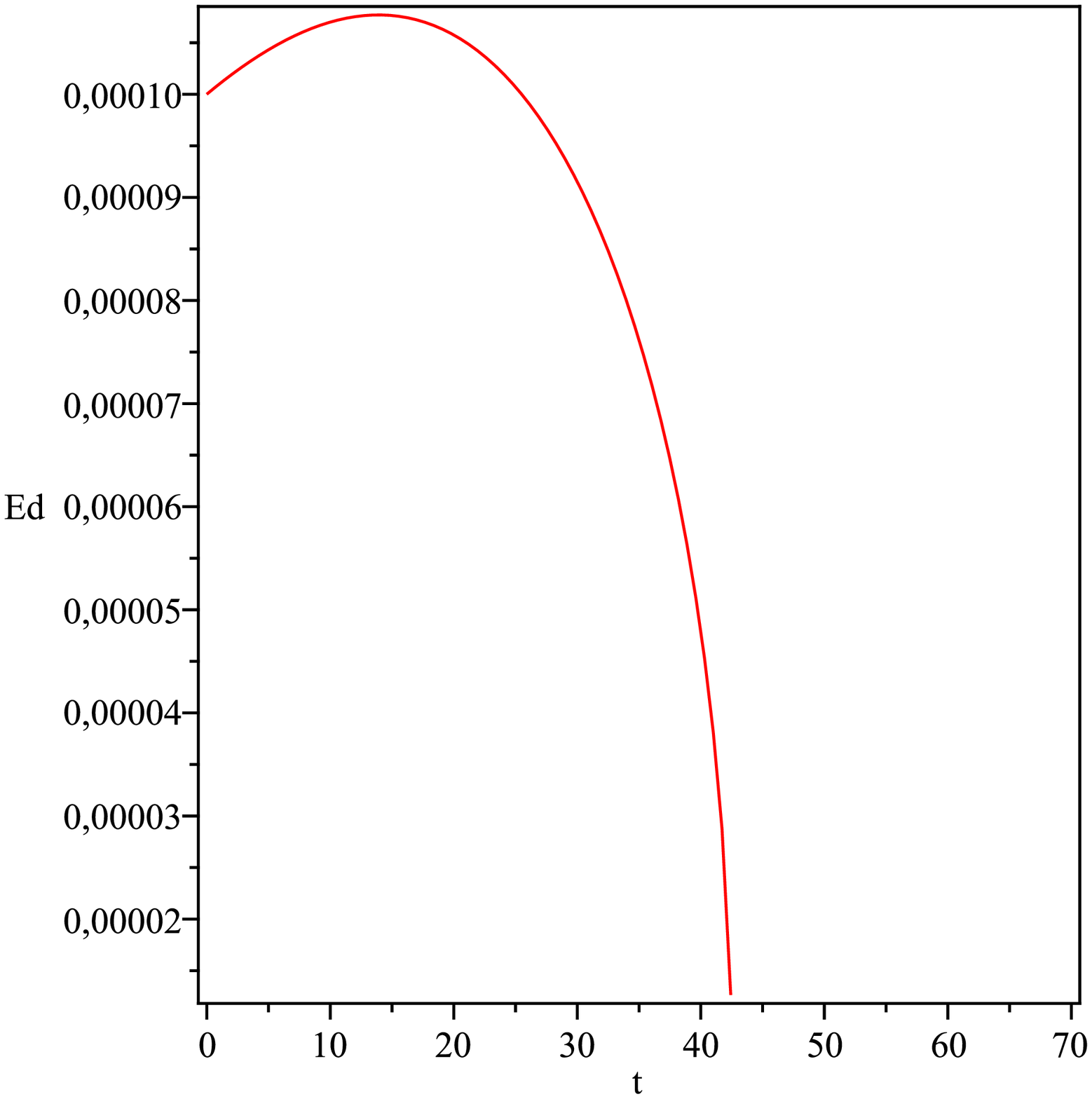}
\caption{$E_d$ for $w=1$.}
\label{fig:2}       % Give a unique label
\end{figure}

\begin{figure}
\includegraphics[scale=0.45]{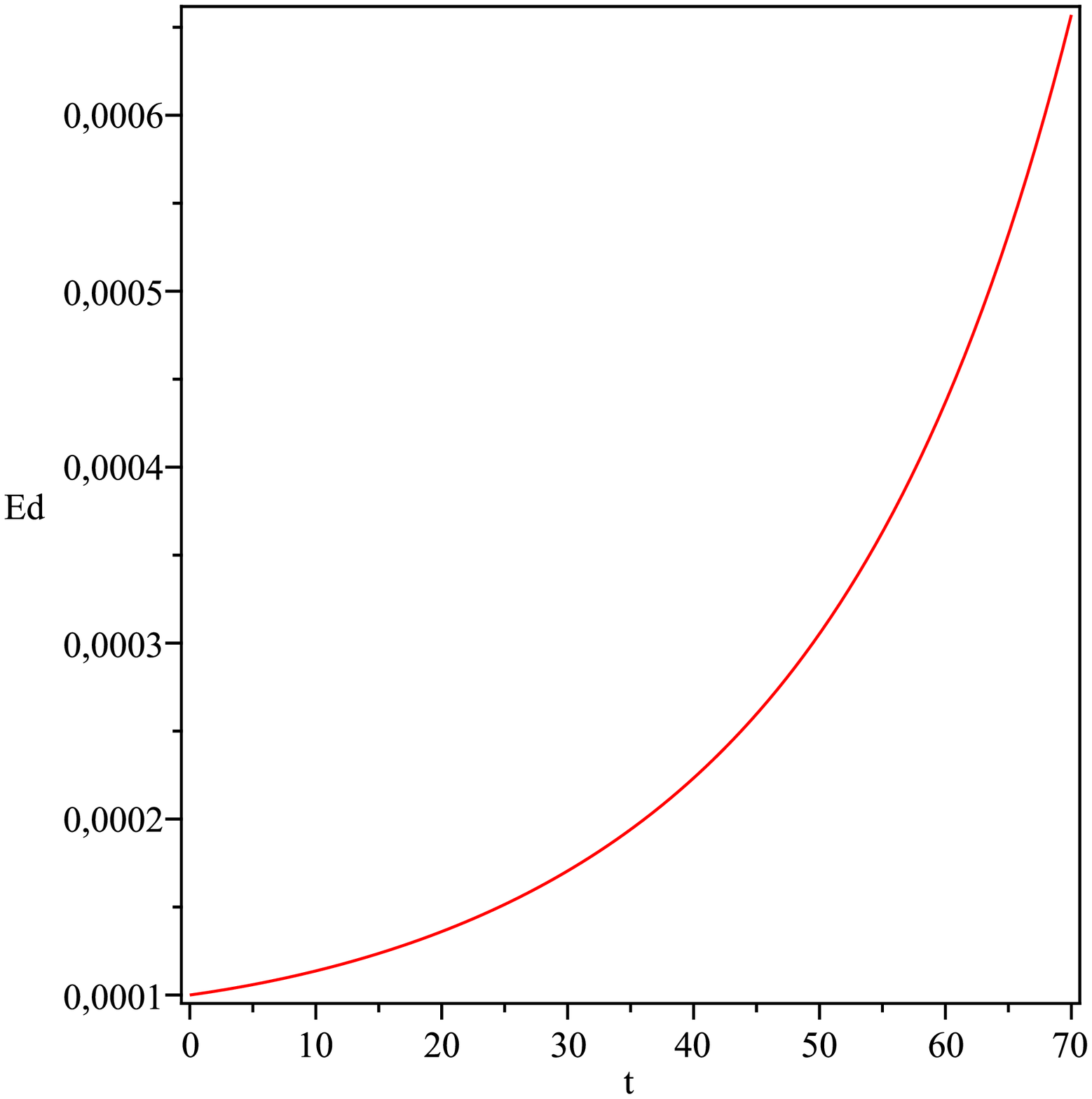}
\caption{$E_d$ for $w=-1$.}
\label{fig:3}       % Give a unique label
\end{figure}

The time dependence of the dark energy is plotted for the following initial conditions: $\epsilon(0)=0.01$, $\dot{\epsilon}(0)=0$, $a(0)=0.01$ and $\phi(0)=0.01$, where $\frac{8\pi\rho}{3}=\epsilon^2$.

\section{Conclusion}

In this article the total energy in conformal teleparallel gravity is defined by a local conservation law which comes from the field equations. In this sense it is possible to tell apart the energy-momentum of the matter fields, the teleparallel energy-momentum and the conformal contribution. Then the total energy and the perfect fluid energy was calculated for the FRW model. As a consequence the so called dark energy can be understood as the difference between those energies since in a vacuum state there is still an energy. From a previous work we obtained the scale factor and the conformal scalar field for a wide range of pertinent parameters and we used them to plot the temporal evolution of the dark energy, for this we adopted the dynamical horizon of the FRW line element. It should be noted that the condition $\kappa=0$ was chosen to fit experimental evidence which indicates that the Universe is approximately flat~\cite{PhysRevLett.111.111302,LHuillier:2016mtc}. It is important to point out that the energy is not static in the FRW Universe, since it depends on the volume of integration. This volume is chosen to be a sphere with the radius equal to the time dependent horizon of the metric which is the radius of the observable Universe. Therefore the bigger the observable Universe the greater is the energy contained in it. Thus it is natural to expect a time dependence for the energy rather than a fixed quantity even if the Universe is closed.

\section*{Acknowledgements}
This work was partially supported by Conselho Nacional de Desenvolvimento Cient\'{i}fico e Tecnol\'{o}gico (CNPq) and Coordena\c{c}\~{a}o de Aperfei\c{c}oamento de Pessoal de N\'{i}vel Superior (CAPES).

%\bibliography{refs}

\begin{thebibliography}{17}%
\makeatletter
\providecommand \@ifxundefined [1]{%
 \@ifx{#1\undefined}
}%
\providecommand \@ifnum [1]{%
 \ifnum #1\expandafter \@firstoftwo
 \else \expandafter \@secondoftwo
 \fi
}%
\providecommand \@ifx [1]{%
 \ifx #1\expandafter \@firstoftwo
 \else \expandafter \@secondoftwo
 \fi
}%
\providecommand \natexlab [1]{#1}%
\providecommand \enquote  [1]{``#1''}%
\providecommand \bibnamefont  [1]{#1}%
\providecommand \bibfnamefont [1]{#1}%
\providecommand \citenamefont [1]{#1}%
\providecommand \href@noop [0]{\@secondoftwo}%
\providecommand \href [0]{\begingroup \@sanitize@url \@href}%
\providecommand \@href[1]{\@@startlink{#1}\@@href}%
\providecommand \@@href[1]{\endgroup#1\@@endlink}%
\providecommand \@sanitize@url [0]{\catcode `\\12\catcode `\$12\catcode
  `\&12\catcode `\#12\catcode `\^12\catcode `\_12\catcode `\%12\relax}%
\providecommand \@@startlink[1]{}%
\providecommand \@@endlink[0]{}%
\providecommand \url  [0]{\begingroup\@sanitize@url \@url }%
\providecommand \@url [1]{\endgroup\@href {#1}{\urlprefix }}%
\providecommand \urlprefix  [0]{URL }%
\providecommand \Eprint [0]{\href }%
\providecommand \doibase [0]{http://dx.doi.org/}%
\providecommand \selectlanguage [0]{\@gobble}%
\providecommand \bibinfo  [0]{\@secondoftwo}%
\providecommand \bibfield  [0]{\@secondoftwo}%
\providecommand \translation [1]{[#1]}%
\providecommand \BibitemOpen [0]{}%
\providecommand \bibitemStop [0]{}%
\providecommand \bibitemNoStop [0]{.\EOS\space}%
\providecommand \EOS [0]{\spacefactor3000\relax}%
\providecommand \BibitemShut  [1]{\csname bibitem#1\endcsname}%
\let\auto@bib@innerbib\@empty
%</preamble>
\bibitem [{\citenamefont {Abbott}\ \emph {et~al.}(2016)\citenamefont {Abbott},
  \citenamefont {Abbott}, \citenamefont {Abbott}, \citenamefont {Abernathy},
  \citenamefont {Acernese}, \citenamefont {Ackley}, \citenamefont {Adams},
  \citenamefont {Adams}, \citenamefont {Addesso}, \citenamefont {Adhikari}
  \emph {et~al.}}]{abbott2016observation}%
  \BibitemOpen
  \bibfield  {author} {\bibinfo {author} {\bibfnamefont {B.}~\bibnamefont
  {Abbott}}, \bibinfo {author} {\bibfnamefont {R.}~\bibnamefont {Abbott}},
  \bibinfo {author} {\bibfnamefont {T.}~\bibnamefont {Abbott}}, \bibinfo
  {author} {\bibfnamefont {M.}~\bibnamefont {Abernathy}}, \bibinfo {author}
  {\bibfnamefont {F.}~\bibnamefont {Acernese}}, \bibinfo {author}
  {\bibfnamefont {K.}~\bibnamefont {Ackley}}, \bibinfo {author} {\bibfnamefont
  {C.}~\bibnamefont {Adams}}, \bibinfo {author} {\bibfnamefont
  {T.}~\bibnamefont {Adams}}, \bibinfo {author} {\bibfnamefont
  {P.}~\bibnamefont {Addesso}}, \bibinfo {author} {\bibfnamefont
  {R.}~\bibnamefont {Adhikari}},  \emph {et~al.},\ }\href@noop {} {\bibfield
  {journal} {\bibinfo  {journal} {Physical review letters}\ }\textbf {\bibinfo
  {volume} {116}},\ \bibinfo {pages} {061102} (\bibinfo {year}
  {2016})}\BibitemShut {NoStop}%
\bibitem [{\citenamefont {Garecki}(2002)}]{garecki2002gravitational}%
  \BibitemOpen
  \bibfield  {author} {\bibinfo {author} {\bibfnamefont {J.}~\bibnamefont
  {Garecki}},\ }\href@noop {} {\bibfield  {journal} {\bibinfo  {journal}
  {Annalen der Physik}\ }\textbf {\bibinfo {volume} {11}},\ \bibinfo {pages}
  {442} (\bibinfo {year} {2002})}\BibitemShut {NoStop}%
\bibitem [{\citenamefont {Kox}\ \emph {et~al.}(1998)\citenamefont {Kox},
  \citenamefont {Klein}, \citenamefont {Schulmann},\ and\ \citenamefont
  {Kilmister}}]{kox1998collected}%
  \BibitemOpen
  \bibfield  {author} {\bibinfo {author} {\bibfnamefont {A.~J.}\ \bibnamefont
  {Kox}}, \bibinfo {author} {\bibfnamefont {M.~J.}\ \bibnamefont {Klein}},
  \bibinfo {author} {\bibfnamefont {R.}~\bibnamefont {Schulmann}}, \ and\
  \bibinfo {author} {\bibfnamefont {C.}~\bibnamefont {Kilmister}},\ }\href@noop
  {} {\  (\bibinfo {year} {1998})}\BibitemShut {NoStop}%
\bibitem [{\citenamefont {Arnowitt}\ \emph {et~al.}(2008)\citenamefont
  {Arnowitt}, \citenamefont {Deser},\ and\ \citenamefont
  {Misner}}]{Arnowitt:1962hi}%
  \BibitemOpen
  \bibfield  {author} {\bibinfo {author} {\bibfnamefont {R.~L.}\ \bibnamefont
  {Arnowitt}}, \bibinfo {author} {\bibfnamefont {S.}~\bibnamefont {Deser}}, \
  and\ \bibinfo {author} {\bibfnamefont {C.~W.}\ \bibnamefont {Misner}},\
  }\href {\doibase 10.1007/s10714-008-0661-1} {\bibfield  {journal} {\bibinfo
  {journal} {Gen. Rel. Grav.}\ }\textbf {\bibinfo {volume} {40}},\ \bibinfo
  {pages} {1997} (\bibinfo {year} {2008})},\ \Eprint
  {http://arxiv.org/abs/gr-qc/0405109} {arXiv:gr-qc/0405109 [gr-qc]}
  \BibitemShut {NoStop}%
%%CITATION = GR-QC/0405109;%%
\bibitem [{\citenamefont {Perlmutter}\ \emph {et~al.}(1999)\citenamefont
  {Perlmutter} \emph {et~al.}}]{Perlmutter:1998np}%
  \BibitemOpen
  \bibfield  {author} {\bibinfo {author} {\bibfnamefont {S.}~\bibnamefont
  {Perlmutter}} \emph {et~al.} (\bibinfo {collaboration} {Supernova Cosmology
  Project}),\ }\href {\doibase 10.1086/307221} {\bibfield  {journal} {\bibinfo
  {journal} {Astrophys. J.}\ }\textbf {\bibinfo {volume} {517}},\ \bibinfo
  {pages} {565} (\bibinfo {year} {1999})},\ \Eprint
  {http://arxiv.org/abs/astro-ph/9812133} {arXiv:astro-ph/9812133} \BibitemShut
  {NoStop}%
%%CITATION = ASTRO-PH/9812133;%%
\bibitem [{\citenamefont {Riess}\ \emph {et~al.}(1998)\citenamefont {Riess}
  \emph {et~al.}}]{Riess:1998cb}%
  \BibitemOpen
  \bibfield  {author} {\bibinfo {author} {\bibfnamefont {A.~G.}\ \bibnamefont
  {Riess}} \emph {et~al.} (\bibinfo {collaboration} {Supernova Search Team}),\
  }\href {\doibase 10.1086/300499} {\bibfield  {journal} {\bibinfo  {journal}
  {Astron. J.}\ }\textbf {\bibinfo {volume} {116}},\ \bibinfo {pages} {1009}
  (\bibinfo {year} {1998})},\ \Eprint {http://arxiv.org/abs/astro-ph/9805201}
  {arXiv:astro-ph/9805201} \BibitemShut {NoStop}%
%%CITATION = ASTRO-PH/9805201;%%
\bibitem [{\citenamefont {{Silvestri}}\ and\ \citenamefont
  {{Trodden}}(2009)}]{Silvestri:2009hh}%
  \BibitemOpen
  \bibfield  {author} {\bibinfo {author} {\bibfnamefont {A.}~\bibnamefont
  {{Silvestri}}}\ and\ \bibinfo {author} {\bibfnamefont {M.}~\bibnamefont
  {{Trodden}}},\ }\href {\doibase 10.1088/0034-4885/72/9/096901} {\bibfield
  {journal} {\bibinfo  {journal} {Reports on Progress in Physics}\ }\textbf
  {\bibinfo {volume} {72}},\ \bibinfo {pages} {096901} (\bibinfo {year}
  {2009})},\ \Eprint {http://arxiv.org/abs/0904.0024} {arXiv:0904.0024
  [astro-ph.CO]} \BibitemShut {NoStop}%
\bibitem [{\citenamefont {Maluf}(2005)}]{maluf2005gravitational}%
  \BibitemOpen
  \bibfield  {author} {\bibinfo {author} {\bibfnamefont {J.}~\bibnamefont
  {Maluf}},\ }\href@noop {} {\bibfield  {journal} {\bibinfo  {journal} {Annalen
  der Physik}\ }\textbf {\bibinfo {volume} {14}},\ \bibinfo {pages} {723}
  (\bibinfo {year} {2005})}\BibitemShut {NoStop}%
\bibitem [{\citenamefont {Maluf}\ and\ \citenamefont
  {Ulhoa}(2008)}]{Maluf:2008yy}%
  \BibitemOpen
  \bibfield  {author} {\bibinfo {author} {\bibfnamefont {J.~W.}\ \bibnamefont
  {Maluf}}\ and\ \bibinfo {author} {\bibfnamefont {S.~C.}\ \bibnamefont
  {Ulhoa}},\ }\href {\doibase 10.1103/PhysRevD.78.069901,
  10.1103/PhysRevD.78.047502} {\bibfield  {journal} {\bibinfo  {journal} {Phys.
  Rev.}\ }\textbf {\bibinfo {volume} {D78}},\ \bibinfo {pages} {047502}
  (\bibinfo {year} {2008})},\ \bibinfo {note} {[Erratum: Phys.
  Rev.D78,069901(2008)]},\ \Eprint {http://arxiv.org/abs/0807.0255}
  {arXiv:0807.0255 [gr-qc]} \BibitemShut {NoStop}%
%%CITATION = ARXIV:0807.0255;%%
\bibitem [{\citenamefont {Ulhoa}\ \emph {et~al.}(2010)\citenamefont {Ulhoa},
  \citenamefont {da~Rocha~Neto},\ and\ \citenamefont
  {Maluf}}]{ulhoa2010gravitational}%
  \BibitemOpen
  \bibfield  {author} {\bibinfo {author} {\bibfnamefont {S.}~\bibnamefont
  {Ulhoa}}, \bibinfo {author} {\bibfnamefont {J.}~\bibnamefont
  {da~Rocha~Neto}}, \ and\ \bibinfo {author} {\bibfnamefont {J.}~\bibnamefont
  {Maluf}},\ }\href@noop {} {\bibfield  {journal} {\bibinfo  {journal}
  {International Journal of Modern Physics D}\ }\textbf {\bibinfo {volume}
  {19}},\ \bibinfo {pages} {1925} (\bibinfo {year} {2010})}\BibitemShut
  {NoStop}%
\bibitem [{\citenamefont {Ulhoa}\ and\ \citenamefont
  {Amorim}(2014)}]{Ulhoa:2014eka}%
  \BibitemOpen
  \bibfield  {author} {\bibinfo {author} {\bibfnamefont {S.~C.}\ \bibnamefont
  {Ulhoa}}\ and\ \bibinfo {author} {\bibfnamefont {R.~G.~G.}\ \bibnamefont
  {Amorim}},\ }\href {\doibase 10.1155/2014/812691} {\bibfield  {journal}
  {\bibinfo  {journal} {Adv. High Energy Phys.}\ }\textbf {\bibinfo {volume}
  {2014}},\ \bibinfo {pages} {812691} (\bibinfo {year} {2014})},\ \Eprint
  {http://arxiv.org/abs/1405.0540} {arXiv:1405.0540 [gr-qc]} \BibitemShut
  {NoStop}%
%%CITATION = ARXIV:1405.0540;%%
\bibitem [{\citenamefont {Maluf}(2013)}]{Maluf:2013gaa}%
  \BibitemOpen
  \bibfield  {author} {\bibinfo {author} {\bibfnamefont {J.~W.}\ \bibnamefont
  {Maluf}},\ }\href {\doibase 10.1002/andp.201200272} {\bibfield  {journal}
  {\bibinfo  {journal} {Annalen Phys.}\ }\textbf {\bibinfo {volume} {525}},\
  \bibinfo {pages} {339} (\bibinfo {year} {2013})},\ \Eprint
  {http://arxiv.org/abs/1303.3897} {arXiv:1303.3897 [gr-qc]} \BibitemShut
  {NoStop}%
%%CITATION = ARXIV:1303.3897;%%
\bibitem [{\citenamefont {Maluf}\ and\ \citenamefont
  {Faria}(2012)}]{Maluf:2011kf}%
  \BibitemOpen
  \bibfield  {author} {\bibinfo {author} {\bibfnamefont {J.~W.}\ \bibnamefont
  {Maluf}}\ and\ \bibinfo {author} {\bibfnamefont {F.~F.}\ \bibnamefont
  {Faria}},\ }\href {\doibase 10.1103/PhysRevD.85.027502} {\bibfield  {journal}
  {\bibinfo  {journal} {Phys. Rev.}\ }\textbf {\bibinfo {volume} {D85}},\
  \bibinfo {pages} {027502} (\bibinfo {year} {2012})},\ \Eprint
  {http://arxiv.org/abs/1110.3095} {arXiv:1110.3095 [gr-qc]} \BibitemShut
  {NoStop}%
%%CITATION = ARXIV:1110.3095;%%
\bibitem [{\citenamefont {{Nesbet}}(2013)}]{2013Entrp..15..162N}%
  \BibitemOpen
  \bibfield  {author} {\bibinfo {author} {\bibfnamefont {R.}~\bibnamefont
  {{Nesbet}}},\ }\href {\doibase 10.3390/e15010162} {\bibfield  {journal}
  {\bibinfo  {journal} {Entropy}\ }\textbf {\bibinfo {volume} {15}},\ \bibinfo
  {pages} {162} (\bibinfo {year} {2013})},\ \Eprint
  {http://arxiv.org/abs/1208.4972} {arXiv:1208.4972 [physics.gen-ph]}
  \BibitemShut {NoStop}%
\bibitem [{\citenamefont {Silva}\ \emph {et~al.}(2016)\citenamefont {Silva},
  \citenamefont {Santos},\ and\ \citenamefont {Ulhoa}}]{silva2016friedmann}%
  \BibitemOpen
  \bibfield  {author} {\bibinfo {author} {\bibfnamefont {J.}~\bibnamefont
  {Silva}}, \bibinfo {author} {\bibfnamefont {A.}~\bibnamefont {Santos}}, \
  and\ \bibinfo {author} {\bibfnamefont {S.}~\bibnamefont {Ulhoa}},\
  }\href@noop {} {\bibfield  {journal} {\bibinfo  {journal} {The European
  Physical Journal C}\ }\textbf {\bibinfo {volume} {76}},\ \bibinfo {pages} {1}
  (\bibinfo {year} {2016})}\BibitemShut {NoStop}%
\bibitem [{\citenamefont {Liddle}\ and\ \citenamefont
  {Cort\^es}(2013)}]{PhysRevLett.111.111302}%
  \BibitemOpen
  \bibfield  {author} {\bibinfo {author} {\bibfnamefont {A.~R.}\ \bibnamefont
  {Liddle}}\ and\ \bibinfo {author} {\bibfnamefont {M.}~\bibnamefont
  {Cort\^es}},\ }\href {\doibase 10.1103/PhysRevLett.111.111302} {\bibfield
  {journal} {\bibinfo  {journal} {Phys. Rev. Lett.}\ }\textbf {\bibinfo
  {volume} {111}},\ \bibinfo {pages} {111302} (\bibinfo {year}
  {2013})}\BibitemShut {NoStop}%
\bibitem [{\citenamefont {L'Huillier}\ and\ \citenamefont
  {Shafieloo}(2017)}]{LHuillier:2016mtc}%
  \BibitemOpen
  \bibfield  {author} {\bibinfo {author} {\bibfnamefont {B.}~\bibnamefont
  {L'Huillier}}\ and\ \bibinfo {author} {\bibfnamefont {A.}~\bibnamefont
  {Shafieloo}},\ }\href {\doibase 10.1088/1475-7516/2017/01/015} {\bibfield
  {journal} {\bibinfo  {journal} {JCAP}\ }\textbf {\bibinfo {volume} {1701}},\
  \bibinfo {pages} {015} (\bibinfo {year} {2017})},\ \Eprint
  {http://arxiv.org/abs/1606.06832} {arXiv:1606.06832 [astro-ph.CO]}
  \BibitemShut {NoStop}%
%%CITATION = ARXIV:1606.06832;%%
\end{thebibliography}
%\bibliographystyle{apsrev4-1}

%

\end{document}